\documentclass[aps,pre,showpacs]{revtex4}

\usepackage{graphicx}

\newcommand{\figurewidth}{0.48\textwidth}

\begin{document}

\title{Generalized Geometric Cluster Algorithm for Fluid Simulation}
\author{Jiwen Liu}
\author{Erik Luijten}
\email[Corresponding author. E-mail: ]{luijten@uiuc.edu}
\affiliation{Department of Materials Science and Engineering and
  Frederick Seitz Materials Research Laboratory, University of
  Illinois at Urbana-Champaign, Urbana, Illinois 61801}

\date{January 17, 2005}

\begin{abstract}
  We present a detailed description of the generalized geometric cluster
  algorithm for the efficient simulation of continuum fluids. The
  connection with well-known cluster algorithms for lattice spin models
  is discussed, and an explicit full cluster decomposition is derived
  for a particle configuration in a fluid. We investigate a number of
  basic properties of the geometric cluster algorithm, including the
  dependence of the cluster-size distribution on density and
  temperature. Practical aspects of its implementation and possible
  extensions are discussed.  The capabilities and efficiency of our
  approach are illustrated by means of two example studies.
\end{abstract}


\maketitle

\section{Introduction}

Computer simulation methods play an increasingly important role in the
study of complex fluids. Historically, many simulations have
concentrated on fluids modeled as an assembly of monodisperse spherical
particles interacting, e.g., via a bare excluded-volume potential
(hard-sphere fluid) or a Lennard-Jones
interaction~\cite{allentildesley87}. While such systems can already
display a wealth of interesting features, such as a solid--liquid
transition and---in the presence of attractive interactions---a critical
point, it is also clear that real fluids and solutions can exhibit far
richer behavior. Various factors contribute to the additional features
found in these systems, including internal degrees of freedom of the
constituents (such as in polymeric systems), interactions induced by
electrical charges, and the presence of various species with a strong
size asymmetry. Modeling any of these properties can greatly increase
the required numerical efforts and in certain situations can make the
simulations prohibitively expensive.  Although available computational
power continues to increase steadily, further progress in simulating
such systems will critically depend on algorithmic advances.

Recently, we have introduced a simulation method that addresses one of
the above-mentioned complicating factors, namely the slow-down arising
in simulations of solutions containing species of largely different
sizes~\cite{geomc} (see also Ref.~\cite{day04}).  This method, which
generalizes an original idea due to Dress and Krauth~\cite{dress95} to
identify clusters based upon geometric symmetry operations and
accordingly is called the (generalized) \emph{geometric cluster
algorithm} (GCA), exhibits two noteworthy features.  Firstly, it employs
a nonlocal Monte Carlo (MC) scheme to move the constituent particles in
a nonphysical way, thus introducing artificial dynamics while preserving
all thermodynamic equilibrium properties.  This greatly accelerates the
generation of uncorrelated configurations of particles.  We emphasize
that our scheme does not involve any approximations; the molecular
configurations produced by the GCA are generated according to the same
Boltzmann distribution that would govern a conventional simulation of
the system under consideration, but the GCA follows a different
trajectory through phase space.  Secondly, the nonlocal moves involve
clusters of particles that are constructed in such a way that \emph{all
proposed moves are accepted}. This is not only an additional factor
contributing to the efficiency of the method, but it also has a more
profound significance. Namely, in order to satisfy detailed
balance~\cite{gubernatis03,landau00,frenkel-smit2,newman99} without
imposing the usual Metropolis~\cite{metropolis53} acceptance criterion,
the ratio of the probability of transforming a particle
configuration~$C$ into a new configuration~$C'$ and the probability of
the reverse transformation must be identical to the ratio of the
Boltzmann factors of the configuration~$C'$ and the original
configuration~$C$.  Clearly, this can only be realized if the
transformations (``moves'') are proposed with a probability that
involves knowledge about the physical properties of a system.  For the
vast majority of MC algorithms, this is not the case. The most
well-known exceptions are the Swendsen--Wang (SW)
algorithm~\cite{swendsen87} for lattice spin models and its variant due
to Wolff~\cite{wolff89}. Among continuum systems, hard-sphere fluids
constitute a special case, since all configurations without particle
overlap have the same energy and thus the same Boltzmann factor.
However, it is generally accepted that these \emph{rejection-free}
algorithms are an exception (cf., e.g., Ref.~\cite{frenkel-smit2},
Sec.~14.3.1).  Indeed, ever since the invention of the SW cluster
algorithm for Ising and Potts models, its extension to fluids has been a
widely-pursued goal.  For lattice gases the extension can be
accomplished in a straightforward manner, since they are isomorphic to
Potts models. However, for off-lattice (continuum) fluids, this mapping
cannot be applied, owing to the absence of particle--hole symmetry.
This symmetry is a critical ingredient for the SW algorithm for Ising
and Potts models, as clusters of spins (or variables, in the case of the
Potts model) are identified using a symmetry operation that, if applied
globally, would leave the Hamiltonian of the system invariant.  The only
known extensions to continuum systems apply to the Widom--Rowlinson
model for fluid mixtures~\cite{johnson97,chayes98}, in which identical
particles do not interact and unlike species experience a hard-core
repulsion, and the closely-related Stillinger--Helfand model, in which
the hard-core is replaced by a soft repulsion~\cite{sun00}. However, no
generalization has been found in which identical particles do not behave
as an ideal gas. In a separate development, Dress and
Krauth~\cite{dress95} observed that, for hard-core fluids, an
overlap-free configuration can be transformed into a new configuration
of nonoverlapping particles by means of a \emph{geometric} symmetry
operation.  The particular advantage of this scheme is its ability to
efficiently relax size-asymmetric
mixtures~\cite{buhot98,santen00,malherbe01}. In Ref.~\cite{geomc} we
showed how this advantage can be retained for fluids with
\emph{arbitrary} pair potentials, while simultaneously exploiting the
invariance of the Hamiltonian under the symmetry operation to create
particle clusters in a manner that is fully analogous to the SW or Wolff
method. An application to a concrete physical problem has been presented
in Ref.~\cite{liu04b}, where colloid--nanoparticle mixtures featuring
attractive, repulsive and hard-core interactions were simulated for size
asymmetries up to~100. We remark that geometric cluster algorithms have
also been applied successfully to lattice-based models. Heringa and
Bl\"ote have employed it to investigate lattice gases with
nearest-neighbor exclusion~\cite{heringa96} and developed a version for
simulation of the Ising model in the constant-magnetization
ensemble~\cite{heringa98}. The ``pocket algorithm'' of Krauth and
Moessner~\cite{krauth03}, which is essentially a geometric cluster
algorithm for dimer models, was used to investigate hard-core dimers on
three-dimensional lattices~\cite{huse03}.  Furthermore, we note that
there is a profound difference between the geometric cluster algorithm
and other Monte Carlo schemes that move groups of particles
simultaneously~\cite{wu92,jaster99,woodcock99,lobaskin99}. Such
algorithms can work efficiently in specific cases, and are sometimes
viewed as counterparts of the SW algorithm merely because they also
invoke the notion of a ``cluster,'' but they typically involve a tunable
parameter and are neither rejection free nor do they exploit a symmetry
property of the Hamiltonian.

In the current article we provide a detailed description of the
generalized geometric cluster algorithm for continuum fluids and study
basic properties such as the dependence of cluster-size distribution on
temperature and density. We also describe a multiple-cluster variant of
the generalized GCA in which a particle configuration is fully
decomposed into clusters that can be moved independently.

\section{Algorithm description}

\subsection{Swendsen--Wang and Wolff algorithms for lattice spin models}
\label{sec:sw-wolff}

For reference in future sections that highlight the similarities between
the generalized GCA~\cite{geomc} for off-lattice fluids and the
SW~\cite{swendsen87} and Wolff~\cite{wolff89} algorithms for lattice
spin models, we briefly summarize the lattice cluster algorithms for the
case of a $d$-dimensional Ising model with nearest-neighbor
interactions, described by the Hamiltonian
\begin{equation}
  \label{eq:ising}
  \mathcal{H}_{\rm Ising} = - J \sum_{\langle ij\rangle} s_i s_j \;.
\end{equation}
The spins $s$ are placed on the vertices of a square ($d=2$) or simple
cubic ($d=3$) lattice and take values $\pm 1$. The sum runs over all
pairs of nearest neighbors, which are coupled via a ferromagnetic
coupling with strength $J>0$. Starting from a given configuration of
spins, the SW algorithm now proceeds as follows:
\begin{enumerate}
\item A ``bond'' is formed between every pair of nearest neighbors that
  are aligned, with a probability $p_{ij} = 1 - \exp(-2 \beta J)$, where
  $\beta = 1/k_{\rm B}T$.
\item All spins that are connected, directly or indirectly, via bonds
  belong to a single cluster. Thus, the bond assignment procedure
  divides the system into clusters of parallel spins. Note how the bond
  probability (and hence the typical cluster size) grows with increasing
  coupling strength~$\beta J$ (decreasing temperature). For finite
  $\beta J$, $p_{ij} < 1$ and hence a cluster is generally a
  \emph{subset} of all spins of a given sign.
\item All spins in each cluster are flipped \emph{collectively} with a
  probability~$\frac{1}{2}$. I.e., for each cluster of spins a spin
  value $\pm 1$ is chosen and this value is assigned to all spins that
  belong to the cluster.
\end{enumerate}
The last, and crucial, step is made possible by the Fortuin--Kasteleyn
mapping~\cite{kasteleyn69,fortuin72} of the Potts model on the
random-cluster model, which implies that the partition function of the
former can be written as a Whitney polynomial that represents the
partition function of the latter~\cite{baxter76}. Accordingly, all spins
in a cluster (a connected component in the random-cluster model) are
uncorrelated with all other spins, and can be assigned a new spin value.

This algorithm possesses several noteworthy properties. First, it
strongly suppresses dynamic slowing down near a critical
point~\cite{swendsen87} by efficiently destroying nonlocal correlations
(see also Ref.~\cite{newman99} for a pedagogical introduction).
Secondly, this algorithm is \emph{rejection free}. Indeed, the
assignment of bonds involves random numbers, but once the clusters have
been formed each of them can be flipped independently without imposing
an acceptance criterion involving the energy change induced by such a
collective spin-reversal operation.  Thirdly, the Fortuin--Kasteleyn
mapping can also be applied to systems in which each spin interacts not
only with its nearest neighbors, but also with other
spins~\cite{baxter76}.  In particular, the coupling strength can be
different for different spin pairs, leading to a probability~$p_{ij}$
that is, e.g., dependent on the separation between $i$ and~$j$.  Thus,
cluster algorithms can be designed for spin systems with
medium-~\cite{medran} and long-range interactions~\cite{lr-alg}.

Critical slowing down is suppressed even more strongly in a variant of
this algorithm due to Wolff~\cite{wolff89}. In this implementation, no
decomposition of the entire spin configuration into clusters takes
place. Instead, a single cluster is formed, which is then always
flipped:
\begin{enumerate}
\item A spin~$i$ is selected at random.
\item All nearest neighbors~$j$ of this spin are added to the cluster
  with a probability $p_{ij} = 1 - \exp(-2 \beta J)$, provided spins $i$
  and~$j$ are parallel and the bond between $i$ and~$j$ has not been
  considered before.
\item Each spin $j$ that is indeed added to the cluster is also placed
  on the stack. Once all neighbors of~$i$ have been considered for
  inclusion in the cluster, a spin is retrieved from the stack and all
  its neighbors are considered in turn for inclusion in the cluster as
  well, following step~2.
\item Steps 2 and~3 are repeated iteratively until the stack is empty.
\item Once the cluster has been completed, all spins in the cluster are
  inverted.
\end{enumerate}
Again, this is a rejection-free algorithm, in the sense that the cluster
is always flipped.  Just as in the SW algorithm, the
cluster-construction process involves random numbers, but the individual
probabilities~$p_{ij}$ involve single-particle energies rather than an
acceptance criterion that involves the total energy change induced by a
cluster flip.

\subsection{Geometric cluster algorithm for hard-sphere mixtures}
\label{sec:hardcore-gca}

Since suppression of critical slowing is a highly attractive feature for
fluid simulations as well, the generalization of the SW and Wolff
algorithms to fluid systems has been a widely-pursued goal. In the
lattice-gas interpretation, where a spin $+1$ corresponds to a particle
and a spin $-1$ corresponds to an empty site, a spin-inversion operation
corresponds to a particle being inserted in or removed from the system.
This ``particle--hole symmetry'' is absent in off-lattice (continuum)
systems. While a particle in a fluid configuration can straightforwardly
be deleted, there is no unambiguous prescription on how to transform
empty space into a particle.  More precisely, in the lattice cluster
algorithms the operation performed on every spin is self-inverse. This
requirement is not fulfilled for off-lattice fluids.

\begin{figure}
\begin{center}
\includegraphics[width=0.96\textwidth]{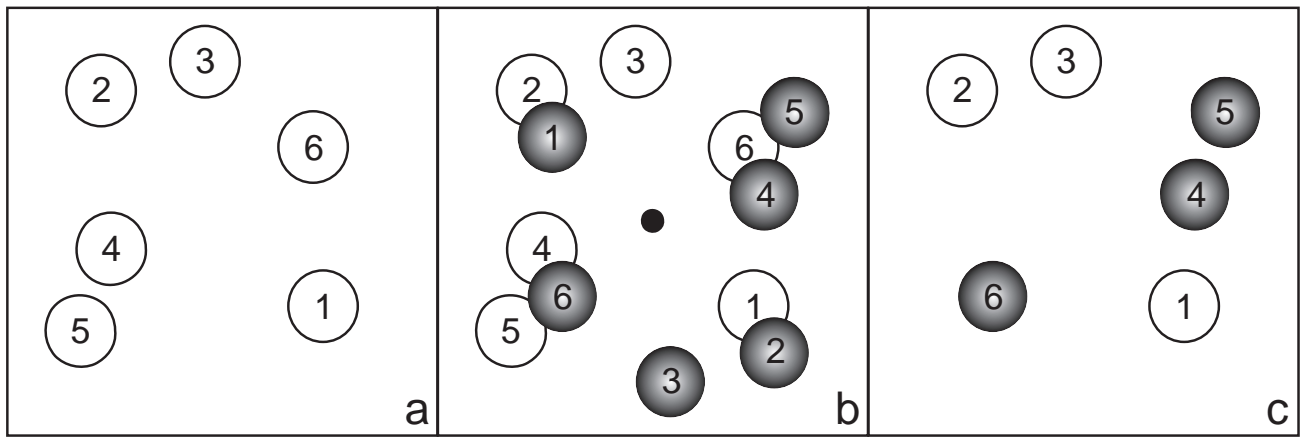}
\caption{Illustration of the geometric cluster algorithm for hard
disks~\protect\cite{dress95}. a)~Original configuration. b)~A new
configuration (shaded circles) is created by means of a point reflection
of all particles with respect to a randomly-chosen pivot point (small
filled disk). The superposition of the original and the new
configuration leads to groups of overlapping particles. In this example,
there are three \emph{pairs} of groups ($\{1,2\}$, $\{3\}$,
$\{4,5,6\}$). Each pair is denoted a \emph{cluster}.  The particles in
any one of these clusters can be point-reflected with respect to the
pivot without affecting the other two clusters. This can be used to
carry out the point reflection for every cluster with a pre-set
probability. c)~Final configuration that results if, starting from the
original configuration, only the particles in the third cluster
$\{4,5,6\}$ are point-reflected.  This approach guarantees that every
generated configuration will be free of overlaps. Note that the pivot
will generally not be placed in the center of the cell, and that the
periodic boundary conditions indeed permit any position.}
\label{fig:geomc-hs}
\end{center}
\end{figure}

Dress and Krauth~\cite{dress95} proposed a method to efficiently
generate particle configurations for a hard-sphere liquid.  This
\emph{geometric cluster algorithm} proceeds as follows (cf.\ 
Fig.~\ref{fig:geomc-hs}).
\begin{enumerate}
\item In a given configuration~$C$, a ``pivot'' is chosen at random.
\item A configuration~$\tilde{C}$ is now generated by carrying out a
  point reflection for all particles in~$C$ with respect to the pivot.
\item The configuration~$C$ and its transformed counterpart~$\tilde{C}$
  are superimposed, which leads to groups of overlapping particles. The
  groups generally come in pairs, except possibly for a single group
  that is symmetric with respect to the pivot. Each pair is denoted a
  ``cluster''~\cite{note-gca}.
\item For each cluster, all particles can be exchanged between $C$
  and~$\tilde{C}$ without affecting particles belonging to other
  clusters. This exchange is performed for each cluster independently
  with a probability~$\smash{\frac{1}{2}}$. Thus, if the superposition
  of $C$ and~$\tilde{C}$ is decomposed into $N$ clusters, there are
  $2^N$ possible new configurations. The configurations that are
  actually realized are denoted~$C'$ and $\tilde{C}'$, i.e., the
  original configuration~$C$ is transformed into $C'$ and its
  point-reflected counterpart~$\tilde{C}$ is transformed
  into~$\tilde{C}'$.
\item The configuration $\tilde{C}'$ is discarded and $C'$ is the new
  configuration, serving as the starting point for the next iteration of
  the algorithm. Note that a new pivot is chosen in every iteration.
\end{enumerate}
Observe that periodic boundary conditions must be employed, such that an
arbitrary placement of the pivot is possible.  Other self-inverse
operations are permissible, such as a reflection in a
plane~\cite{heringa96}, in which case various orientations of the plane
must be chosen in order to satisfy ergodicity.  While operating in the
canonical rather than in the grand-canonical ensemble, this prescription
clearly bears great resemblance to the original SW algorithm.  The
original configuration is decomposed into clusters by exploiting a
symmetry operation that leaves the Hamiltonian invariant if applied to
the entire configuration; in the SW algorithm this is the spin-inversion
operation and in the geometric cluster algorithm it is a geometric
symmetry operation. Subsequently, a new configuration is created by
moving each cluster independently with a certain probability.

We note that this method represents an approach of great generality. For
example, it is not restricted to monodisperse systems, and has indeed
been applied successfully to binary~\cite{buhot98} and
polydisperse~\cite{santen00} mixtures. Indeed, the nonlocal character of
the particle moves makes them exquisitely suitable to overcome the
jamming problems that slow down the simulation of size-asymmetric
mixtures.  An important limitation of the algorithm is the fact that the
average cluster size increases very rapidly beyond a certain density,
corresponding to the percolation threshold of the combined system
containing the superposition of the configurations $C$ and~$\tilde{C}$.
Once this cluster spans the entire system, the algorithm is clearly no
longer ergodic.

We can take the analogy with the lattice cluster algorithms one step
further, by showing that a single-cluster (Wolff) variant can be
formulated as well~\cite{heringa96,geomc}.
\begin{enumerate}
\item In a given configuration~$C$, a ``pivot'' is chosen at random.
\item A particle~$i$ is selected as the first particle that belongs to
  the cluster. This particle is moved via a point reflection with
  respect to the pivot. In its new position, the particle is referred to
  as~$i'$.
\item Step 2 is repeated \emph{iteratively} for each particle~$j$ that
  overlaps with~$i'$. Thus, if the (moved) particle~$j'$ overlaps with
  another particle~$k$, particle~$k$ is moved as well. Note that all
  translations involve the same pivot.
\item Once all overlaps have been resolved, the cluster move is
  completed.
\end{enumerate}
As in the SW-like prescription, a new pivot is chosen for each cluster
that is constructed.

\subsection{Geometric cluster algorithm for interacting particles:
Single-cluster variant}
\label{sec:gca}

The geometric cluster algorithm described in the previous section is
formulated for hard-core interactions. For application to general pair
potentials, it was suggested in Ref.~\cite{dress95} to impose a
Metropolis-type acceptance criterion based upon the energy difference
induced by the cluster move.  Indeed, in the case of potentials that
consist of a hard-core (excluded-volume) contribution supplemented by an
attractive or repulsive tail, such as a Yukawa potential, the
cluster-construction procedure guarantees that no overlaps are generated
and the acceptance criterion takes into account the tail of the
interactions. For ``soft-core'' potentials, such as a Lennard-Jones
interaction, the situation becomes already somewhat more complicated,
since an arbitrary excluded-volume distance must be chosen for the
cluster construction. As the algorithm will not generate configurations
in which the separation between a pair of particles is less than this
distance (i.e., the particle ``diameter,'' in the case of monodisperse
systems), it must be set to a value that is smaller than any separation
that would typically occur, as already noted in Ref.~\cite{malherbe99}.
In either case, the clusters that are generated have only limited
physical relevance, and the evaluation of a considerable part of the
energy change resulting from a cluster move is deferred until the
acceptance step.  Rejection is not only likely, but also costly, given
the computational efforts of both cluster construction and energy
evaluation. We note that this approach was nevertheless applied to
Yukawa mixtures with moderate size asymmetry (diameter ratio $\leq
5$)~\cite{malherbe99}.

On the other hand, Heringa and Bl\"ote~\cite{heringa98,heringa98c}
devised an geometric cluster algorithm for the Ising model in which the
nearest-neighbor interactions between spins are taken into account
already during the cluster construction. While this lattice model can
obviously be simulated by the SW and Wolff algorithms, their approach
permits simulation in the constant-magnetization ensemble. The geometric
operations employed map the spin lattice onto itself, such that
excluded-volume conditions are satisfied automatically: every spin move
amounts to an \emph{exchange} of spins. For every spin pair $(i,i')$
that is exchanged, each of its nearest-neighbor pairs $(k,k')$ are
exchanged with a probability that depends on the change in pair energy,
$\Delta = (E_{ik} + E_{i'k'}) - (E_{ik'} + E_{i'k})$. This procedure is
then again performed iteratively for the neighbors of all spin pairs
that are exchanged.

\begin{figure}
\begin{center}
\includegraphics[width=0.96\textwidth]{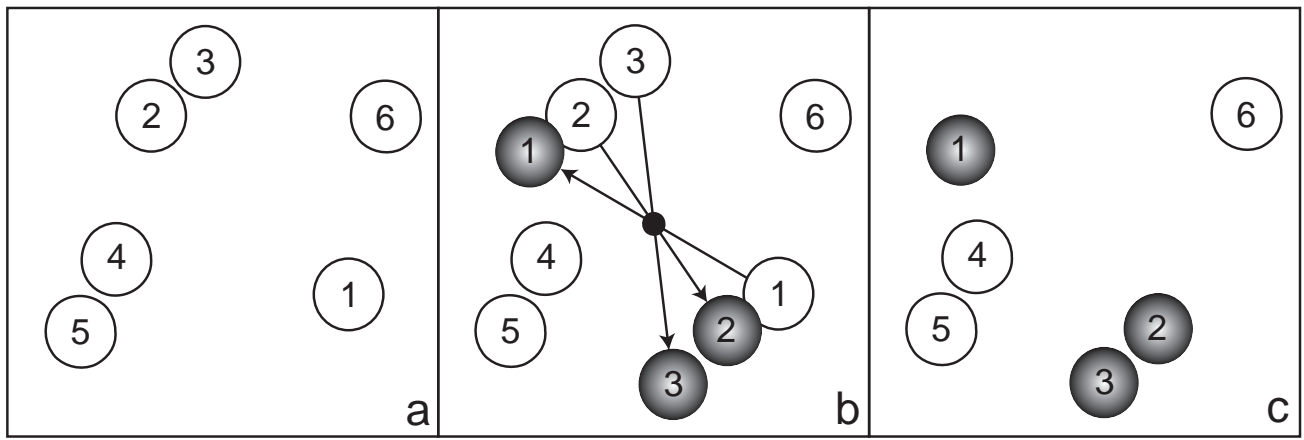}
\caption{Two-dimensional illustration of the interacting geometric
cluster algorithm. Like in Fig.~\protect\ref{fig:geomc-hs}, open and
shaded disks denote the particles before and after the geometric
operation, respectively, and the small disk denotes the pivot. However,
in the \emph{generalized} GCA a single cluster is constructed, to which
particles are added with an interaction-dependent probability.
a)~Original configuration. b)~A cluster is constructed as follows.
Particle~1 is point-reflected with respect to the pivot. If, in its new
position, it has a repulsive interaction with particle~2, the latter has
a certain probability to be point-reflected as well, with respect to the
same pivot. Assuming an attractive interaction between particles 2
and~3, particle~3 is translated as well, but again only with a certain
probability.  If particles 4--6 are not affected by these point
reflections, the cluster construction terminates. c)~The new
configuration consists of particles 1--3 in their new positions and
particles 4--6 in the original positions. A new pivot is chosen and the
procedure is repeated.}
\label{fig:geomc-wolff}
\end{center}
\end{figure}

In Ref.~\cite{geomc} we introduced a generalization of the GCA for
off-lattice fluids, in which particles undergo a geometric operation in
a stochastic manner, akin to the approach of Ref.~\cite{heringa98}. The
differences arise from the fact that no underlying lattice is present,
so that particles are added to the cluster on an individual basis,
rather than in pairs. All interactions are treated in a unified manner,
so there is no technical distinction between attractive and repulsive
interactions or between hard-core and soft-core potentials. This
\emph{generalized GCA} is most easily described as a combination of the
single-cluster methods formulated in Sec.\ \ref{sec:sw-wolff}
and~\ref{sec:hardcore-gca}. We assume a general pair potential
$V_{ij}(\mathbf{r}_{ij})$ that does not have to be identical for all
pairs $(i,j)$ (see Fig.~\ref{fig:geomc-wolff}).
\begin{enumerate}
\item In a given configuration~$C$, a ``pivot'' is chosen at random.
\item A particle~$i$ at position~$\mathbf{r}_i$ is selected as the first
  particle that belongs to the cluster. This particle is moved via a
  point reflection with respect to the pivot. In its new position, the
  particle is referred to as~$i'$, at position~$\mathbf{r}_i'$.
\item Each particle~$j$ that interacts with $i$ or~$i'$ is now
  considered for addition to the cluster. Unlike the first particle,
  particle~$j$ is point-reflected with respect to the pivot only with a
  probability $p_{ij} = \max[1 - \exp(-\beta \Delta_{ij}), 0]$, where
  $\Delta_{ij} = V(|\mathbf{r}_i'-\mathbf{r}_j|) -
  V(|\mathbf{r}_i-\mathbf{r}_j|)$.  A particle~$j$ that interacts with
  $i$ both in its old and in its new position is nevertheless treated
  only once.
\item Each particle~$j$ that is indeed added to the cluster (i.e.,
  moved) is also placed on the stack. Once all particles interacting
  with $i$ or~$i'$ have been considered, a particle is retrieved from
  the stack and all its neighbors that are not yet part of the cluster
  are considered in turn for inclusion in the cluster as well, following
  step~3.
\item Steps 3 and~4 are repeated iteratively until the stack is empty.
  The cluster move is now complete.
\end{enumerate}
If a particle interacts with multiple other particles that have been
added to the cluster, it can thus be considered multiple times for
inclusion. However, once it has been added to the cluster, it cannot be
removed. This is an important point in practice, since particles undergo
a point reflection already during the cluster construction process (and
thus need to be tagged, in order to prevent them from being returned to
their original position by a second point reflection).  A crucial aspect
is that the probability $p_{ij}$ \emph{only} depends on the change in
\emph{pair energy} between $i$ and~$j$. A change in the relative
position of $i$ and~$j$ occurs if particle~$j$ is \emph{not} added to
the cluster. This happens with a probability $1-p_{ij} = \min[
\exp(-\beta \Delta_{ij}), 1]$. The similarity with the Metropolis
acceptance criterion is deceptive (and merely reflects the fact that
both algorithms aim to generate configurations according to the
Boltzmann distribution) , since $\Delta_{ij}$ does not represent the
\emph{total} energy change resulting from the translation of
particle~$i$. Instead, other energy changes are taken into account via
the iterative nature of the algorithm.

It is instructive to note that the expression for $p_{ij}$ bears close
resemblance to the probability employed in the SW algorithm
(Sec.~\ref{sec:sw-wolff}). In the latter, two different situations can
be discerned that lead to a change in the relative energy
$\smash{\Delta_{ij}^{\rm SW}}$ between a spin~$i$ that belongs to the
cluster and a spin~$j$ that does not yet belong to the cluster. If $i$
and $j$ are initially \emph{antiparallel}, $j$ will never be added to
the cluster and only spin~$i$ will be inverted, yielding an energy
change $\smash{\Delta_{ij}^{\rm SW}} = -2J < 0$ that occurs with
probability unity. If $i$ and $j$ are initially \emph{parallel} and $j$
is not added to the cluster, the resulting change in the pair energy
equals $\smash{\Delta_{ij}^{\rm SW}} = +2J > 0$.  This occurs with a
probability $\exp(-2\beta J) < 1$. These two situations can indeed be
summarized as $\min[ \exp(-\beta \Delta_{ij}^{\rm SW}), 1]$, just as in
the generalized GCA\@.

The ergodicity of this algorithm follows from the fact that there is a
nonvanishing probability that a cluster consists of only one particle,
which can be moved over an arbitrarily small distance, since the
location of the pivot is chosen at random. This obviously requires that
not all particles are part of the cluster, a condition that is violated
at high packing fractions. While a compact proof of detailed balance of
the generalized GCA has already been given in Ref.~\cite{geomc}, we
include it here for the sake of completeness. We consider a
configuration~$X$ that is transformed into a configuration~$Y$ by means
of a cluster move.  All particles included in the cluster maintain their
relative separation; as noted above, an energy change arises if a
particle is \emph{not} included in the cluster, but interacts with a
particle that does belong to the cluster.  Following
Wolff~\cite{wolff89} we denote each of these interactions as a ``broken
bond.'' A broken bond~$k$ that corresponds to an energy
change~$\Delta_k$ occurs with a probability $1-p_k = 1$ if $\Delta_k
\leq 0$ and a probability $1 - p_k = \exp(-\beta \Delta_{k})$ if
$\Delta_k > 0$.  The formation of an entire cluster corresponds to the
breaking of a set~$\{k\}$ of bonds, which has a probability~$P$. This
set is comprised of the subset $\{l\}$ of broken bonds~$l$ that lead to
an increase in pair energy and the subset $\{m\}$ of broken bonds that
lead to a decrease in pair energy, such that
\begin{equation}
  P = \prod_{k} (1-p_k) = \exp\left[-\beta \sum_{l} \Delta_l\right] \;.
\end{equation}
The transition probability from configuration~$X$ to configuration~$Y$
is proportional to the cluster formation probability,
\begin{equation}
\label{eq:trans-fw}
T(X\to Y) = C \exp\left[-\beta \sum_{l} \Delta_l\right] \;,
\end{equation}
where the factor $C$ accounts for the fact that various arrangements of
bonds within the cluster (``internal bonds'') correspond to the same set
of broken bonds. In addition, it incorporates the probability of
choosing a particular pivot and a specific particle as the starting
point for the cluster.

If we now consider the reverse transition $Y\to X$, we observe that this
again involves the set~$\{k\}$, but all the energy differences change
sign compared to the forward move. Consequently, the subset $\{l\}$ in
Eq.~(\ref{eq:trans-fw}) is replaced by its complement~$\{m\}$ and the
transition probability is given by
\begin{equation}
\label{eq:trans-rv}
T(Y\to X) = C \exp\left[+ \beta \sum_{m} \Delta_m\right] \;,
\end{equation}
where the factor~$C$ is identical to the prefactor in
Eq.~(\ref{eq:trans-fw}). Since we require the geometric operation to be
self-inverse we thus find that the cluster move satisfies detailed
balance at an acceptance ratio of unity,
\begin{equation}
  \frac{T(X\to Y)}{T(Y\to X)} =
  \frac{\exp\left[-\beta \sum_{l} \Delta_l\right]}%
       {\exp\left[+ \beta \sum_{m} \Delta_m\right]} = \exp\left[ -\beta
  \sum_{k} \Delta_k\right] = \exp\left[ -\beta (E_Y - E_X)\right] =
  \frac{\exp(-\beta E_{Y})}{\exp(-\beta E_{X})} \;,
\end{equation}
where $E_X$ and $E_Y$ are the internal energies of configurations $X$
and~$Y$, respectively. I.e., the ratio of the forward and reverse
transition probabilities is equal to the inverse ratio of the Boltzmann
factors, so that we indeed have created a rejection-free algorithm. This
is obscured to some extent by the fact that in our prescription the
cluster is moved while it is being constructed, similar to the Wolff
algorithm in Sec.~\ref{sec:sw-wolff}. The central point, however, is
that the construction solely involves single-particle energies, whereas
a Metropolis-type approach only evaluates the total energy change
induced by a multi-particle move and then frequently rejects this move.
By contrast, the GCA avoids large energy differences by incorporating
``offending'' particles into the cluster with a high probability.

\subsection{Geometric cluster algorithm for interacting particles:
Full cluster decomposition}
\label{sec:gca-sw}

We now introduce a SW implementation of the generalized GCA\@.  The
merit of this formulation, which builds upon the Wolff version described
in the previous section, is twofold: First, it demonstrates that the
algorithm constitutes the true off-lattice counterpart of the SW and
Wolff cluster algorithms for spin models outlined in
Sec.~\ref{sec:sw-wolff}.  Secondly, the SW formulation produces a full
decomposition of an off-lattice fluid configuration into
\emph{stochastically independent clusters}.  This implies an interesting
and remarkable analogy with the Ising model.  As observed by Coniglio
and Klein~\cite{coniglio80} for the two-dimensional Ising model at its
critical point, the clusters created according to the prescription in
Sec.~\ref{sec:sw-wolff} are just the so-called ``Fisher
droplets''~\cite{fisher67}. While Ref.~\cite{coniglio80} makes no
reference to the work by Fortuin and Kasteleyn, these ``Coniglio--Klein
clusters'' are implied by the Fortuin--Kasteleyn mapping of the Potts
model onto the random-cluster model~\cite{fortuin72}, which in turn
constitutes the basis for the Swendsen--Wang approach~\cite{swendsen87}.
The clusters generated by the GCA do not have an immediate physical
interpretation, as they typically consist of two spatially disconnected
parts. However, just like the Ising clusters can be inverted at random,
each cluster of fluid particles can be moved independently with respect
to the remainder of the system.  As such, the generalized GCA can be
viewed as a continuum version of the Fortuin--Kasteleyn mapping.

The cluster decomposition of a configuration proceeds as follows. First,
a cluster is constructed according to the Wolff version of
Sec.~\ref{sec:gca}, with the exception that the cluster is only
\emph{identified}; particles belonging to the cluster are marked but not
actually moved.  The pivot employed will also be used for the
construction of all subsequent clusters in this decomposition.  These
subsequent clusters are built just like the first cluster, except that
particles that are already part of an earlier cluster will never be
considered for a new cluster. Once each particle is part of exactly one
cluster the decomposition is completed. Like in the SW algorithm, every
cluster can then be moved (i.e., all particles belonging to it are
translated via a point reflection) independently, e.g., with a
probability~$f$. Despite the fact that all clusters except the first are
built in a restricted fashion, each individual cluster is constructed
according to the rules of the Wolff formulation of Sec.~\ref{sec:gca}.
The exclusion of particles that are already part of another cluster
simply corresponds to the fact that every bond should be considered only
once. If a bond is broken during the construction of an earlier cluster
it should not be re-established during the construction of a subsequent
cluster. The cluster decomposition thus obtained is not unique, as it
depends on the placement of the pivot and the choice of the first
particle.  Evidently, this also holds for the SW algorithm.

In order to establish that this prescription is a true equivalent of the
SW algorithm, we prove that each cluster can be moved (reflected)
independently while preserving detailed balance. If only a single
cluster is actually moved, this essentially corresponds to the Wolff
version of the GCA, since each cluster is built according to the GCA
prescription.  The same holds true if several clusters are moved and no
interactions are present between particles that belong to different
clusters (the hard-sphere algorithm is a particular realization of this
situation). If two or more clusters are moved and \emph{broken} bonds
exist between these clusters, i.e., a nonvanishing interaction exists
between particles that belong to disparate (moving) clusters, then the
shared broken bonds are actually preserved and the proof of detailed
balance provided in the previous section no longer applies in its
original form.  However, since these bonds are identical in the forward
and the reverse move, the corresponding factors cancel out. This is
illustrated for the situation of two clusters whose construction
involves, respectively, two sets of broken bonds $\{k_1\}$ and
$\{k_2\}$. Each set comprises bonds $l$ ($\{l_1\}$ and $\{l_2\}$,
respectively) that lead to an \emph{increase} in pair energy and bonds
$m$ ($\{m_1\}$ and $\{m_2\}$, respectively) that lead to a
\emph{decrease} in pair energy. We further subdivide these sets into
\emph{external} bonds that connect cluster 1 or~2 with the remainder of
the system and \emph{joint} bonds that connect cluster~1 with cluster~2.
Accordingly, the probability of creating cluster~1 is given by
\begin{equation}
  C_1 \prod_{i \in \left\{k_1\right\}}(1-p_{i}) = C_1 \prod_{i \in
  \left\{l_1\right\}}(1-p_i) = C_1 \prod_{i \in \left\{l_1^{\rm
  ext}\right\}} (1-p_i) \prod_{j \in \left\{l_1^{\rm joint}\right\}}
  (1-p_j) \;.
\end{equation}
Upon construction of the first cluster, the creation of the second
cluster has a probability
\begin{equation}
  C_2 \prod_{i \in \left\{l_2^{\rm ext}\right\}} (1-p_i) \;,
\end{equation}
since all joint bonds in $\{l_2^{\rm joint}\} = \{l_1^{\rm joint}\}$
already have been broken.  The factors $C_1$ and $C_2$ refer to the
probability of realizing a particular arrangement of internal bonds in
clusters 1 and~2, respectively (cf.\ Sec.~\ref{sec:gca}).  Hence, the
total transition probability of moving \emph{both} clusters is given by
\begin{equation}
  T_{12}(X\to Y) = C_1 C_2 \exp\left[ -\beta \sum_{i \in \left\{l_1^{\rm
  ext}\right\}} \Delta_i -\beta \sum_{j \in \left\{l_2^{\rm
  ext}\right\}} \Delta_j -\beta \sum_{n \in \left\{l_1^{\rm
  joint}\right\}} \Delta_n \right] \;.
\end{equation}
In the reverse move, the energy differences for all external broken
bonds have changed sign, but the energy differences for the joint bonds
connecting cluster 1 and~2 are the same as in the forward move. Thus,
cluster~1 is created with probability
\begin{equation}
  C_1 \prod_{i \in \left\{m_1^{\rm ext}\right\}} (1-\bar{p}_i) \prod_{j
      \in \left\{l_1^{\rm joint}\right\}} (1-p_j) = C_1 \prod_{i \in
      \left\{m_1^{\rm ext}\right\}} \exp[ +\beta \Delta_i ] \prod_{j \in
      \left\{l_1^{\rm joint}\right\}} \exp[-\beta \Delta_j ] \;,
\end{equation}
where the $\bar{p}_i$ reflects the sign change of the energy differences
compared to the forward move and the product over the external bonds
involves the complement of the set $\{l_1^{\rm ext}\}$. The creation
probability for the second cluster is
\begin{equation}
  C_2 \prod_{i \in \left\{m_2^{\rm ext}\right\}} (1-\bar{p}_i) = C_2
  \prod_{i \in \left\{m_2^{\rm ext}\right\}} \exp[ +\beta \Delta_i ]
\end{equation}
and the total transition probability for the reverse move is
\begin{equation}
  T_{12}(Y\to X) = C_1 C_2 \exp\left[ + \beta \sum_{i \in \left\{m_1^{\rm
  ext}\right\}} \Delta_i + \beta \sum_{j \in \left\{m_2^{\rm
  ext}\right\}} \Delta_j -\beta \sum_{n \in \left\{l_1^{\rm
  joint}\right\}} \Delta_n \right] \;.
\end{equation}
Accordingly, detailed balance is still fulfilled with an acceptance
ratio of unity,
\begin{equation}
  \frac{T_{12}(X\to Y)}{T_{12}(Y\to X)} = \exp\left[ -\beta \sum_{i \in
  \left\{k_1^{\rm ext}\right\}} \Delta_i -\beta \sum_{j \in
  \left\{k_2^{\rm ext}\right\}} \Delta_j \right] = \exp\left[ - \beta
  (E_Y - E_X ) \right] \;,
\end{equation}
in which $\{k_1^{\rm ext}\} = \left\{l_1^{\rm ext}\right\} \cup
\left\{m_1^{\rm ext}\right\}$ and $\{k_2^{\rm ext}\} = \left\{l_2^{\rm
  ext}\right\} \cup \left\{m_2^{\rm ext}\right\}$ and $E_X$ and $E_Y$
refer to the internal energy of the system before and after the move,
respectively.  This treatment applies to any simultaneous move of
clusters, so that \emph{each cluster in the decomposition indeed can be
moved independently} without violating detailed balance. This completes
the proof of the multiple-cluster version of the GCA\@. It is noteworthy
that the probabilities for breaking joint bonds in the forward and
reverse moves cancel only because the probability in the cluster
construction factorizes into individual probabilities.

\begin{figure}
\begin{center}
\includegraphics[width=\figurewidth]{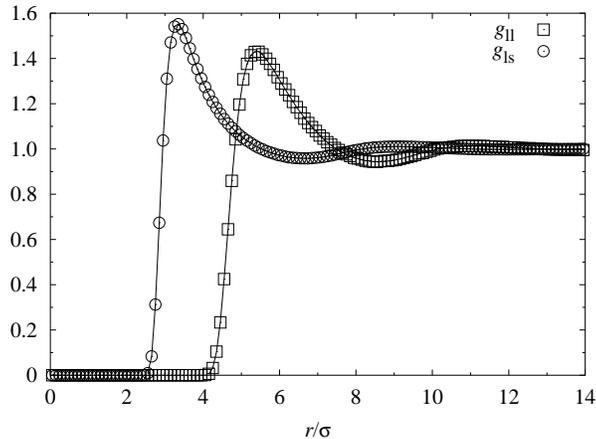}
\caption{Comparison between the single-cluster version
(Sec.~\protect\ref{sec:gca}; solid lines) and the multiple-cluster
version (Sec.~\protect\ref{sec:gca-sw}; symbols) of the generalized
geometric cluster algorithm. The figure shows pair correlation functions
for the size-asymmetric Lennard-Jones mixture described in the text.
$g_{\textrm{ll}}$ and $g_{\textrm{ls}}$ represent the large--large and
large--small correlation functions, respectively. There is excellent
agreement between both algorithms.}
\label{fig:wolff-sw}
\end{center}
\end{figure}

In order to illustrate the validity of this approach, we have applied it
to the binary Lennard-Jones mixture employed in Ref.~\cite{geomc}. This
system consists of $800$ small particles (diameter $\sigma_{11}=1.0$;
reduced coupling $\beta\varepsilon_{11} = 0.40$) and $400$ large
particles (diameter $\sigma_{22}=5.0$; reduced coupling
$\beta\varepsilon_{22} = 0.225$) at a total packing fraction $\phi
\approx 0.213$. Following the Lorentz--Berthelot mixing
rules~\cite{allentildesley87} we set the parameters for the large--small
Lennard-Jones interaction to $\sigma_{12} = (\sigma_{11}+\sigma_{22})/2$
and $\varepsilon_{12} = \sqrt{\varepsilon_{11} \varepsilon_{22}}$.  The
particles are contained in a cubic cell with linear size $L=50$.
Periodic boundary conditions are applied and the cut-off for all
interactions is set to~$3\sigma_{22}$.  We perform a full cluster
decomposition for every configuration and carry out a reflection for
every cluster with a probability $f=\frac{1}{2}$.  As illustrated in
Fig.~\ref{fig:wolff-sw}, the correlation functions for pairs of large
particles and for pairs of large and small particles agree perfectly
with the results obtained in Ref.~\cite{geomc} by means of the
single-cluster version. In Sec.~\ref{sec:efficiency} we address the
relative efficiency of both approaches.

\subsection{Implementation issues}

The actual implementation of the generalized GCA involves a variety of
issues. The point reflection with respect to the pivot requires careful
consideration of the periodic boundary conditions. Furthermore, as
mentioned above, particles that have been translated via a point
reflection must not be translated again in the same cluster move, and
particles that interact with a given cluster particle both before and
after the translation of that cluster particle must be considered only
once, on the basis of the difference in pair potential. In order to
account for all interacting pairs in an efficient manner, we employ the
cell index method~\cite{allentildesley87}. For mixtures with large size
asymmetries (the situation where the generalized GCA excels), it is
natural to set up different cell structures, with cell lengths based
upon the cut-offs of the various particle interactions. For example, in
the case of a binary mixture of two species with very different sizes
and cutoff radii ($\smash{r_{\rm cut}^{\rm large}}$ and $r_{\rm
cut}^{\rm small}$, respectively), the use of an identical cell structure
with a cell size that is determined by the large particles would be
highly inefficient for the smaller particles. Thus, two cell structures
are constructed in this case (with cell sizes $l_{\rm large}$
and~$l_{\rm small}$, respectively) and each particle is stored in the
appropriate cell of the structure belonging to its species, and
incorporated in the corresponding linked list, following the standard
approach~\cite{allentildesley87}. However, in order to efficiently deal
with interactions between unlike species (which have a cut-off~$r_{\rm
cut}^{\rm ls}$), a mapping between the two cell structures is required.
If all small particles must be located that interact with a given large
particle, we proceed as follows. First, the small cell~$\mathbf{c}$ is
identified in which the center of the large particle resides.
Subsequently, the interacting particles are located by scanning over all
small cells within a cubic box with linear size~$2r_{\rm cut}^{\rm ls}$,
centered around~$\mathbf{c}$. This set of cells is predetermined at the
beginning of a run and their indices are stored in an array. Each set
contains approximately $N_{\rm cell}=(2\smash{r_{\rm cut}^{\rm
ls}}/l_{\rm small})^3$ members.  In an efficient implementation, $l_{\rm
small}$ is not much larger than $r_{\rm cut}^{\rm small}$, which for
short-range interactions is of the order of the size of a small
particle. Likewise, $r_{\rm cut}^{\rm ls}$ is typically of the order of
the size of the large particle, so that $N_{\rm cell} =
\mathcal{O}(\alpha^3)$, where $\alpha$ denotes the size asymmetry
between the two species. Since $N_{\rm cell}$ indices must be stored for
each large cell, the memory requirements become very large for cases
with large size asymmetry, cf.\ Ref.~\cite{liu04b} for $\alpha=100$.

\section{Algorithm properties}

\subsection{Efficiency}
\label{sec:efficiency}

\begin{figure}
\begin{center}
\includegraphics[width=\figurewidth]{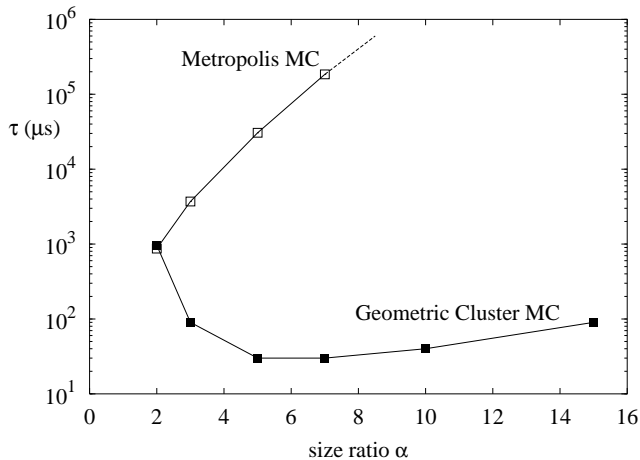}
\caption{Efficiency comparison between a conventional local update algorithm
(open symbols) and the generalized geometric cluster algorithm (closed
symbols), for a binary mixture (see text) with size ratio~$\alpha$.
Whereas the autocorrelation time per particle (expressed in $\mu$s of
CPU time per particle move) rapidly increases with size ratio, the GCA
features only a weak dependence on~$\alpha$.}
\label{fig:efficiency}
\end{center}
\end{figure}

The most notable feature of the generalized GCA, as emphasized in
Ref.~\cite{geomc}, is the efficiency with which it generates
uncorrelated configurations for size-asymmetric mixtures. This
performance directly derives from the nonlocal character of the point
reflection employed. In general, the translation of a single particle
over large distances is impossible in all but the most dilute
situations.  On the other hand, multiple-particle moves typically entail
an energy difference that strongly suppresses the likelihood of
acceptance. By enabling collective moves while maintaining a high (and,
in fact, maximal) acceptance probability, fluid mixtures can be
simulated efficiently over a wide parameter range (volume fraction, size
asymmetry and temperature). Following Ref.~\cite{geomc}, we illustrate
this for a simple binary mixture in which the autocorrelation time is
determined as a function of size asymmetry. This system contains 150
large particles of size $\sigma_{22}$, at fixed volume fraction $\phi_2
= 0.1$.  Furthermore, $N_1$ small particles are present, also at fixed
volume fraction $\phi_1 = 0.1$. Thus, as the size~$\sigma_{11}$ of these
small particles is varied from $\sigma_{22}/2$ to $\sigma_{22}/15$
(i.e., the size ratio $\alpha = \sigma_{22}/\sigma_{11}$ is increased
from 2 to~15), their number increases from $N_1 = 1\,200$ to~$506\,250$.
Pairs of small particles and pairs involving a large and a small
particle act like hard spheres.  However, in order to prevent
depletion-driven aggregation of the large particles~\cite{asakura54}, we
introduce a short-ranged Yukawa repulsion,
\begin{equation}
\label{eq:yu}
U_{ll}(r) = \left \{ \begin{array}{ll} +\infty & \quad r \le
  \sigma_{22}\\ \frac{\sigma_{22}}{r}\varepsilon \exp[-\kappa
  (r-\sigma_{22})] & \quad r > \sigma_{22}
  \end{array} \right.   \;,
\end{equation}
where $\beta\varepsilon = 3.0$ and the screening length $\kappa^{-1} =
\sigma_{11}$.  In the simulation, the exponential tail is cut off at
$3\sigma_{\rm 22}$.

The additional Yukawa interactions also provides a fluctuating internal
energy~$E(t)$ that permits us to determine the rate at which the system
decorrelates. We consider the integrated autocorrelation time~$\tau$
obtained from the energy autocorrelation function~\cite{binder01},
\begin{equation}
\label{eq:correlation}
C(t) =
\frac{\langle{E(0)E(t)}\rangle -   {\langle{E(0)}\rangle}^2}%
      {\langle{E(0)^2}\rangle - {\langle{E(0)}\rangle}^2} \;,
\end{equation}
and compare $\tau$ for a conventional (Metropolis) MC algorithm and the
generalized GCA\@.  In order to avoid arbitrariness resulting from the
computational cost involved with a single sweep or the construction of a
cluster, we express $\tau$ in actual CPU time (assuming that both
methodologies have been programmed in an efficient manner).
Furthermore, $\tau$ is normalized by the total number of particles in
the system, to account for the variation in~$N_1$ as the size ratio
$\alpha$ is increased.

For conventional MC calculations, $\tau_{\rm MC}$ rapidly increases with
increasing~$\alpha$, because the large particles tend to get trapped by
the small particles. Indeed, already for $\alpha > 7$ it is not feasible
to obtain an accurate estimate for~$\tau_{\rm MC}$. By contrast,
$\tau_{\rm GCA}$ exhibits a very different dependence on~$\alpha$. At
$\alpha=2$ both algorithms require virtually identical simulation time,
which establishes that the GCA does not involve considerable overhead
compared to standard algorithms (if any, it is mitigated by the fact
that all moves are accepted). Upon increase of $\alpha$, $\tau_{\rm
GCA}$ initially \emph{decreases} until it starts to increase weakly. The
nonmonotonic variation of $\tau_{\rm GCA}$ results from the changing
ratio $N_2/N_1$ which causes the cluster composition to vary
with~$\alpha$. The main points to note are: (i)~the GCA greatly
suppresses the autocorrelation time, $\tau_{\rm GCA} \ll \tau_{\rm MC}$
for $\alpha > 2$, with an efficiency increase that amounts to more than
three orders of magnitude already for $\alpha = 7$; (ii)~the increase of
the autocorrelation time with $\alpha$ is much slower for the GCA than
for a local-move MC algorithm, making the GCA increasingly advantageous
with increasing size asymmetry. This second observation is confirmed by
the large size asymmetries that could be attained in Ref.~\cite{liu04b}.

\subsection{Cluster size}

\begin{figure}
\includegraphics[width=\figurewidth]{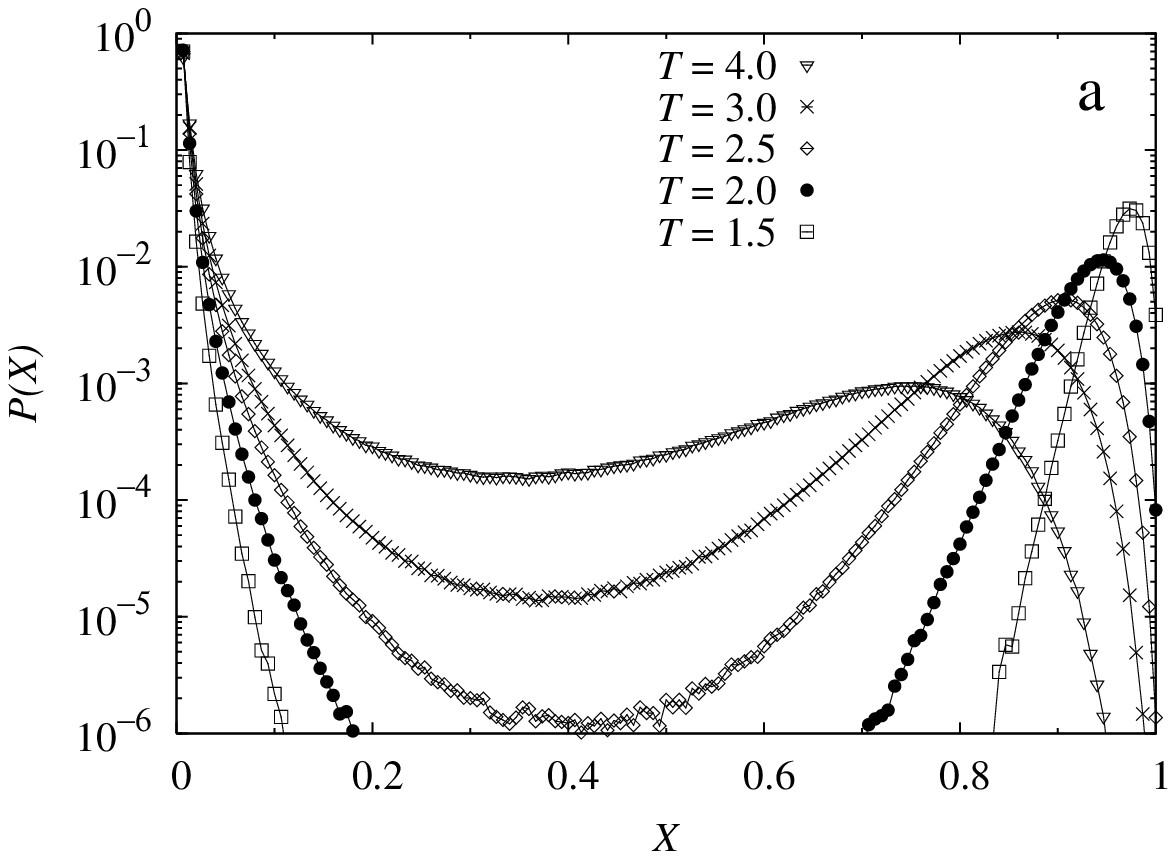}
\includegraphics[width=\figurewidth]{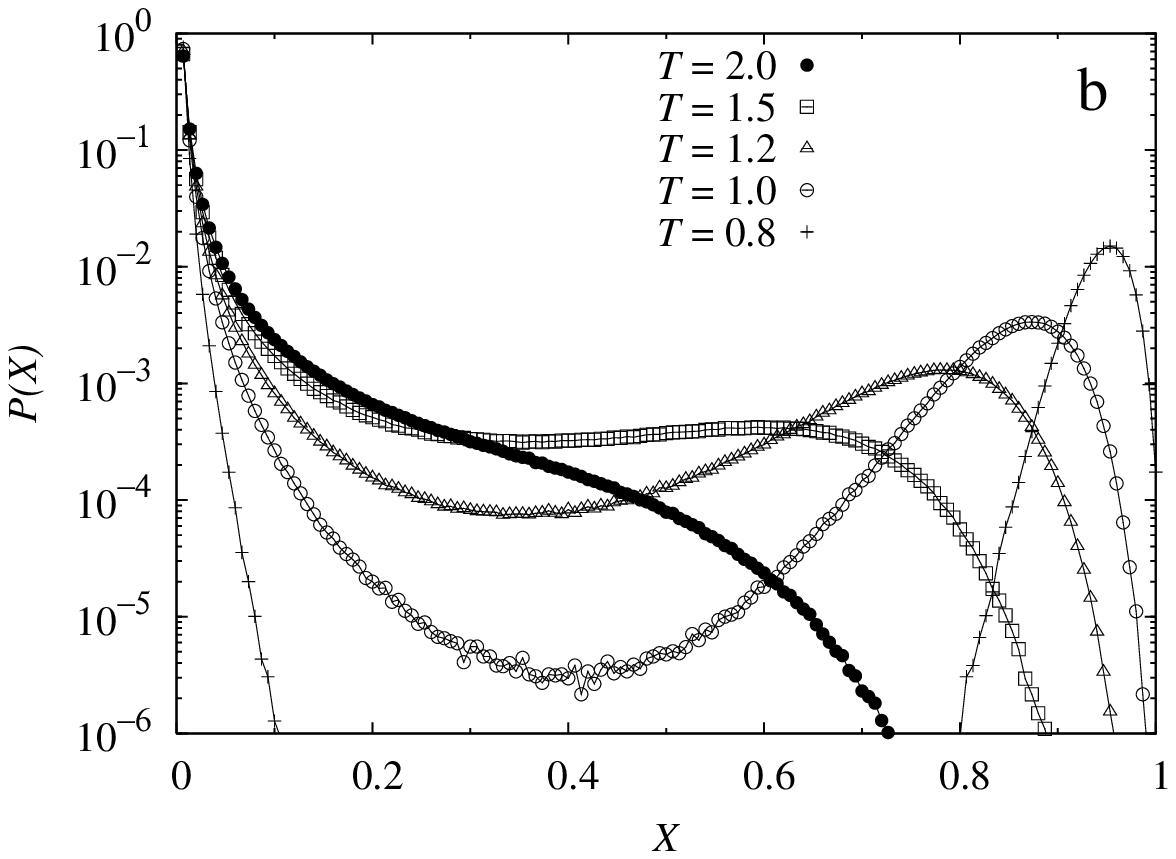}
\caption{Cluster-size distributions as a function of relative cluster
size~$X$, for a monodisperse Lennard-Jones fluid. (a)~Critical density,
$\rho = 0.32\sigma^{-3}$. The distribution is strongly bimodal in the
vicinity of the critical temperature and remains bimodal up to
relatively high temperatures.  (b)~Twice smaller density $\rho =
0.16\sigma^{-3}$. The cluster-size distribution only becomes bimodal for
temperatures relatively close to the critical temperature and decreases
monotonously for higher temperatures.  Identical symbols refer to
identical temperatures in both panels. All temperatures are indicated in
terms of $\varepsilon/k_{\rm B}$.}
\label{fig:clusterdistr}
\end{figure}

The cluster size clearly has a crucial influence on the performance of
the GCA\@. If a cluster contains more than 50\% of all particles, an
equivalent change to the system could have been made by moving its
complement; unfortunately it is unclear how to determine this complement
without constructing the cluster. Nevertheless, it is found that the
algorithm operates in a comparatively efficient manner for average
relative cluster sizes as large as 90\% or more. Once the total packing
fraction of the system exceeds a certain value, the original hard-core
GCA breaks down because each cluster occupies the entire system. The
same phenomenon occurs in the generalized GCA, but in addition the
cluster size can saturate because of strong interactions. Thus, the
maximum accessible volume fraction depends on a considerable number of
parameters, including the range of the potentials and the temperature.
For multi-component mixtures, size asymmetry and relative abundance of
the components are of importance as well, and the situation can be
complicated further by the presence of competing interactions.

As an illustration, we consider the cluster-size distribution for a
monodisperse Lennard-Jones fluid (particle diameter~$\sigma$). For an
interaction cut-off of $2.5\sigma$, the critical temperature~$T_{\rm c}$
lies just below $1.19\varepsilon/k_{\rm B}$~\cite{wilding95}, where
$\varepsilon$ denotes the coupling strength.
Figure~\ref{fig:clusterdistr}(a) shows this distribution at the critical
density~$0.32\sigma^{-3}$ ($\phi \approx 0.168$) for a range of
supercritical temperatures. Already at temperatures that are far above
the critical temperature, the cluster-size distribution tends toward a
bimodal form, indicative of the formation of large clusters.  The gap
between the two peaks widens with decreasing temperature and in the
vicinity of the critical temperature the average cluster size becomes
very large.  This is greatly different from the SW and Wolff algorithms,
which operate at the percolation threshold when applied to a critical
system.  Remarkably, when applied to a lattice gas at its critical
density, the geometric cluster algorithm was also found to yield the
power-law distribution that is characteristic for a percolating
system~\cite{heringa98b}. This can be understood from the fact that
excluded-volume effects play no role in geometric operations applied to
lattice-based systems. In continuum systems, the superposition of a
system and its point-reflected counterpart percolates already when the
original system has a density that is considerably below the percolation
threshold.  Motivated by this, we investigate the cluster-size
distribution in the same Lennard-Jones fluid at a twice lower density,
$\rho = 0.16 \sigma^{-3}$, see Fig.~\ref{fig:clusterdistr}(b). For the
highest temperature (which is already twice lower than the highest
temperature in panel (a)), this distribution now is a monotonously
decreasing function of cluster size, and the bimodal character only
appears for temperatures around 25\% above the critical temperature.

It turns out to be possible to influence the cluster-size distribution
by placing the pivot in a biased manner. Rather than first choosing the
pivot location, a particle is selected that will become the first member
of the cluster. Subsequently, the pivot is placed at random within a
cubic box of linear size~$\delta>0$, centered around the position of
this particle. By decreasing~$\delta$, the displacement of the first
particle is decreased, as well as the number of other particles affected
by this displacement. As a consequence, the average cluster size
decreases, and higher volume fractions can be reached. Ultimately, the
cluster size will still occupy the entire system, but we found that the
maximum accessible volume fraction can be increased from approximately
$0.23$ to a value close to~$0.34$. This value indeed corresponds to the
percolation threshold for hard spheres, $0.3419$~\cite{lorenz01}.  Note
that the proof of detailed balance is not affected by this modification.

\subsection{Critical slowing down}

\begin{figure}
\begin{center}
\includegraphics[width=\figurewidth]{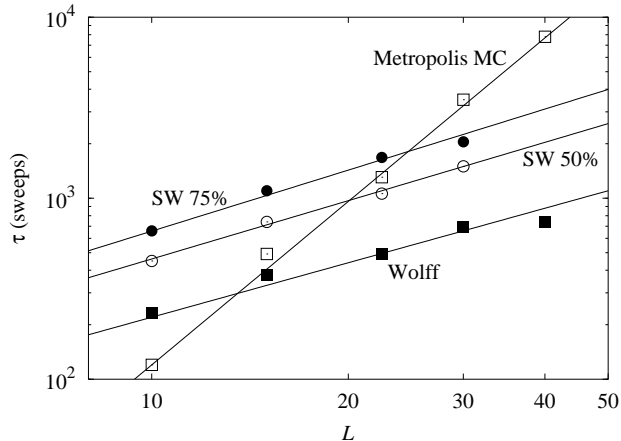}
\caption{Energy autocorrelation time $\tau$ as a function of linear
system size for the critical Lennard-Jones fluid, in units of particle
sweeps, for three different Monte Carlo algorithms: Local moves
(``Metropolis MC''); GCA with Swendsen--Wang type cluster decomposition
and probability 0.50 (``SW 50\%'') and 0.75 (``SW 75\%'') of moving each
cluster; single-cluster GCA (``Wolff'').}
\label{fig:critdyn}
\end{center}
\end{figure}
 
The cluster-size distributions obtained in the previous section suggest
that the generalized GCA will not suppress critical slowing down for the
Lennard-Jones fluid. As emphasized in Ref.~\cite{geomc}, this does not
have to be viewed as a great shortcoming of the algorithm, because of
the efficiency improvement it delivers for the simulation of
size-asymmetric fluids over a wide range of temperatures and packing
fractions (cf.\ Fig.~\ref{fig:efficiency}). Nevertheless, since
suppression of critical slowing down plays such an important role for
lattice cluster algorithms and since it is a feature that has not been
realized by any fluid simulation algorithm, we have investigated the
integrated autocorrelation time for the energy at the critical point, as
a function of linear system size.  In Fig.~\ref{fig:critdyn} these times
are collected for three algorithms.  (1)~Conventional local-update
Metropolis algorithm; (2)~Wolff version of the GCA; (3a)~SW version of
the GCA, in which each cluster is point-reflected with a
probability~$0.50$; (3b)~SW version of the GCA, in which each cluster is
point-reflected with a probability~$0.75$. This also serves as a
performance comparison between the single-cluster GCA and the
multiple-cluster variant.  Just as for spin models, the single-cluster
version is more efficient than the SW-like approach.  However, all
variants of the GCA exhibit the same power-law behavior, which
outperforms the Metropolis algorithm by a factor $\sim L^{2.1}$.  It is
important to emphasize that this acceleration may be due to the
suppression of the hydrodynamic slowing down~\cite{hohenberg77} caused
by the conservation of the density (which may couple to the energy
correlations~\cite{geomc}). Remarkably, already for moderate system
sizes the generalized GCA outperforms the Metropolis algorithm, despite
the time-consuming construction of large clusters [cf.\ 
Fig.~\ref{fig:clusterdistr}(a)] which lead to only small configurational
changes.

\section{Examples}

\subsection{Size-asymmetric mixture with Yukawa attraction between unlike
species}

As an illustration of the capabilities of the generalized GCA we apply
it to a binary mixture first studied in Ref.~\cite{malherbe99}.  It
contains dilute colloidal particles in a solvent of smaller
particles. Both species are modeled as hard spheres, but unlike pairs
experience a Yukawa attraction which promotes the accumulation of the
solvent particles around the colloids.  Systems like this are relevant
for an improved understanding of depletion effects in the presence of
additional nonadditive interactions~\cite{louis01b}, but have hitherto
been studied only to a limited extent, because of computational
limitations.

Specifically, we set up a simulation cell containing 29\,000 small
particles (species~1) at volume fraction~$\phi_1 = 0.116$ and two large
colloids (species~2) at volume fraction~$\phi_2 = 0.001$. The pair
potentials are chosen as
\begin{equation}
  V_{ii} = \left\{
    \begin{array}{ll}
      +\infty & \quad r \leq \sigma_{ii} \\ 0 & \quad r > \sigma_{ii}
    \end{array}
    \right.  \quad i = 1,2
\end{equation}
and
\begin{equation}
  V_{12} = \left\{
    \begin{array}{ll}
      +\infty & \quad r \leq \sigma_{12} \\ -\frac{\sigma_{11}}{r}
      \varepsilon\exp\left[-\kappa(r-\sigma_{12})\right] & \quad r >
      \sigma_{12}
    \end{array}
    \right. \;,
\end{equation}
where $\sigma_{12} = (\sigma_{11}+\sigma_{22})/2$ and
$\sigma_{22}/\sigma_{11} = 5$. The coupling strength is set to
$\varepsilon = 1.6 k_{\rm B}T$, corresponding to a contact energy
$(1.6/3) k_{\rm B}T$, and the decay parameter (inverse screening length)
is set to $\kappa=4/\sigma_{11}$~\cite{note-parameters}.

\begin{figure}
\begin{center}
\includegraphics[width=\figurewidth]{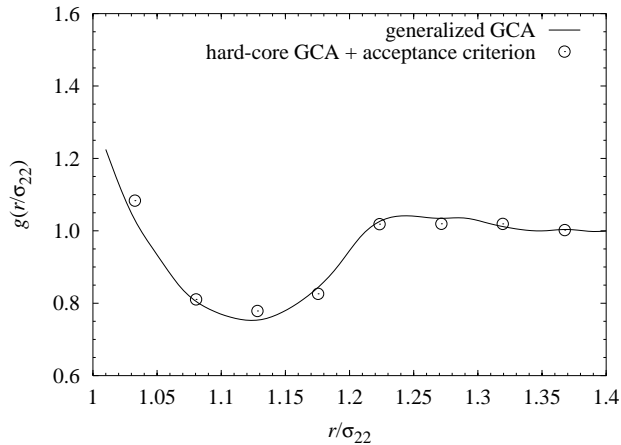}
\caption{Pair correlation function of dilute colloidal particles
(diameter~$\sigma_{22}$ and volume fraction~$\phi_2 = 0.001$) in an
environment of smaller particles (diameter~$\sigma_{11}=\sigma_{22}/5$
and volume fraction~$\phi_1 = 0.116$) that experience a Yukawa-type
attractive interaction with the colloids.  The symbols represent data
obtained by means of the hard-core GCA supplemented by an acceptance
criterion~\protect\cite{malherbe99}. The solid line is obtained via the
generalized GCA.}
\label{fig:amokrane}
\end{center}
\end{figure}

In Ref.~\cite{malherbe99} this system was investigated by means of the
hard-core GCA supplemented by an acceptance criterion in order to take
into account the Yukawa attractions. In view of the size asymmetry
between the two species, this already yields a considerable efficiency
improvement over conventional MC simulations that only employ local
moves. However, the acceptance criterion potentially greatly
deteriorates performance, as clusters will be constructed that are
subsequently rejected in their entirety. Indeed, the authors
report~\cite{malherbe99} that accurate direct sampling of the colloidal
pair correlation function $g(r)$ was prohibitively expensive, so that
they instead obtained it via numerical differentiation of the
\emph{integrated} pair correlation function (see symbols in
Fig.~\ref{fig:amokrane}). This differentiation involves a polynomial fit
and the result was found to be sensitive to the degree of this
polynomial.

In the generalized GCA the Yukawa attractions are directly incorporated
in the cluster construction, so that all clusters are accepted. As
demonstrated in Fig.~\ref{fig:amokrane} (solid line), accurate results
for $g(r)$ can now be obtained through direct sampling. Note that our
choice for the colloid concentration~$\phi_2$ is slightly smaller than
in Ref.~\cite{malherbe99} (leading to a somewhat larger number of small
particles in our calculation), which is however irrelevant for the
results, as they have already converged to the dilute colloid limit.
Thus, the generalized GCA opens possibilities for a systematic
investigation of the effect of interaction strength and range on the
potential of mean force, as a function of size asymmetry and solvent
concentration.

\subsection{Entropic interactions induced by ``soft'' depletants}

The addition of small, nonadsorbing additives, e.g., polymers, to a
colloidal suspension can lead to the well-known depletion interaction
between colloids. This interaction has been modeled successfully by the
Asakura--Oosawa~(AO) model~\cite{asakura54,vrij76} which treats the
polymer chains as ideal, noninteracting spheres that have an
excluded-volume interaction with the colloids. Alternatively, the
polymers can be treated as hard spheres, leading to an additive binary
hard-sphere mixture~\cite{roth00}.  Although on a qualitative level both
theories agree with the experimentally observed trends for depletion
interactions, it recently has been suggested that a more accurate
description can be obtained by means of a model in which the polymer
pair potential is described by a
Gaussian~\cite{louis00a,louis00b,likos01},
\begin{equation}
\label{eq:gauss}
V(r) = \varepsilon \exp\left[ - \left(\frac{r}{R}\right)^2 \right] \;,
\end{equation}
where $\varepsilon$ is the strength of the repulsive potential and $R =
\alpha R_{\rm G}$ its width. For dilute and semidilute polymer
solutions, $\alpha = 1.13$ and $1.45$, respectively, were found to be
appropriate parameter values~\cite{louis00b}. This potential can be
viewed as a model that interpolates between the AO model for ideal
polymers and the binary hard-sphere mixture.

\begin{figure}
\includegraphics[width=\figurewidth]{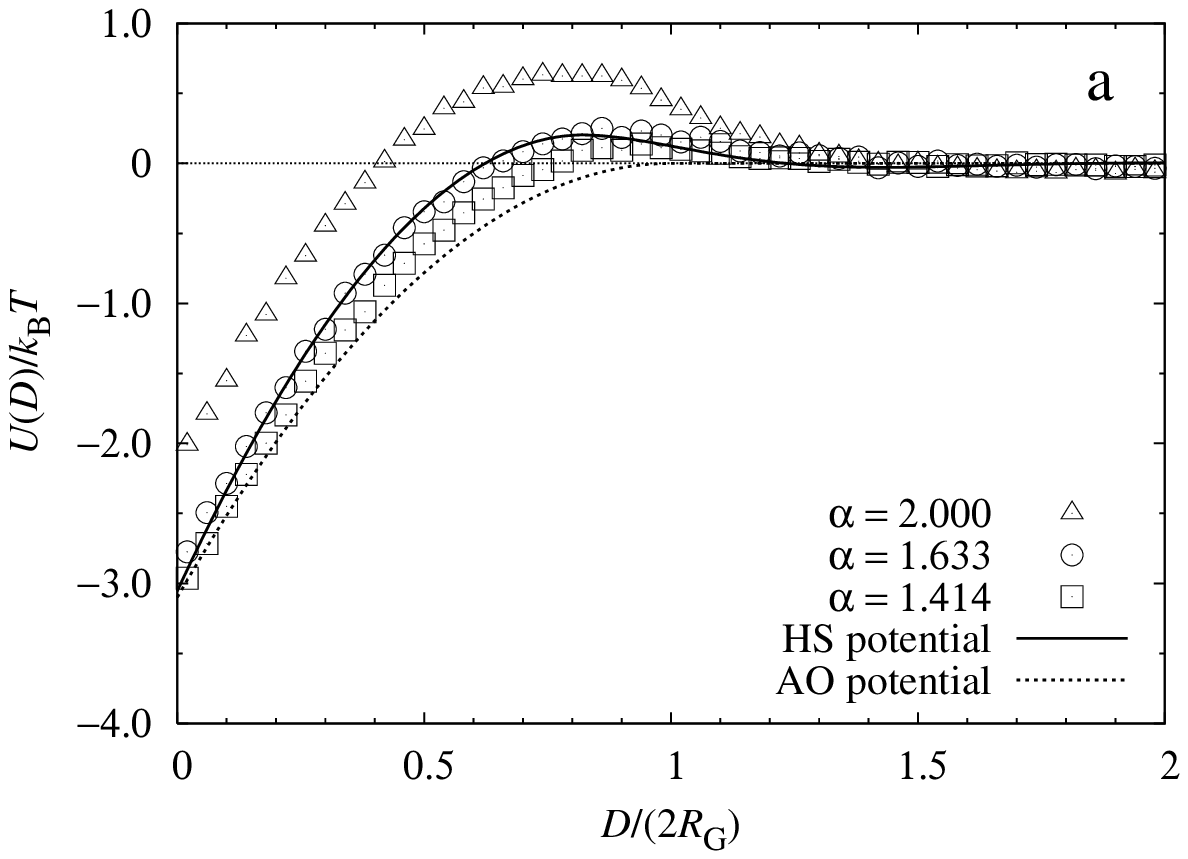}
\includegraphics[width=\figurewidth]{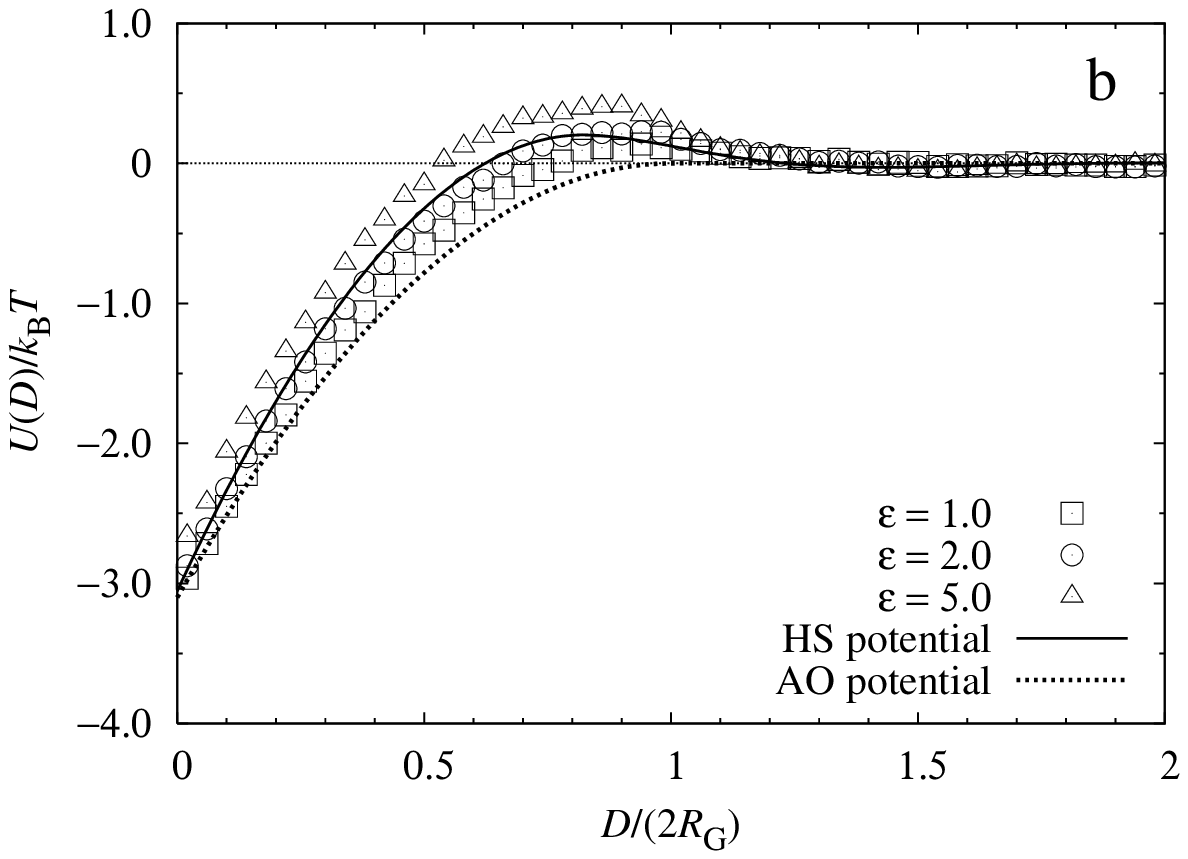}
\caption{Effective pair potential as a function of colloid
surface-to-surface separation~$D$ for the Gaussian polymer model,
Eq.~(\protect\ref{eq:gauss}).  Panel (a) pertains to a fixed interaction
strength $\varepsilon = 1.0 k_{\rm{B}}T$ for various interaction widths
$\alpha = 2.0$, $1.633$, $1.414$ (triangles, circles and squares,
respectively). In panel (b), $\varepsilon$ is varied ($1.0k_{\rm{B}}T$,
$2.0k_{\rm{B}}T$ and~$5.0k_{\rm{B}}T$, indicated by squares, circles and
triangles, respectively) for fixed $\alpha = 1.414$.  The solid line in
both figures represents the hard-sphere result obtained by
density-functional theory~\protect\cite{roth00} and the dotted line
pertains to the expression by Vrij~\protect\cite{vrij76} for the AO
model~\protect\cite{asakura54}.}
\label{fig:gauss}
\end{figure}

In order to demonstrate that these system can be accurately and
efficiently simulated by means of the generalized GCA, we study the
depletion interactions between colloidal particles as a function of the
width~$\alpha$ and the strength $\varepsilon$. The colloid and polymer
volume fractions are fixed at $0.010$ and~$0.10$, respectively, and
their size ratio is set to~$20$. We employ the polymer coil diameter
($2R_{\rm G}$) as unit length scale.  The simulation involves $1.6
\times 10^6$ ``polymers'' and 20 colloidal particles.
Figure~\ref{fig:gauss} shows that the AO model generally overestimates
the depletion attraction in both strength and range. Indeed, the
soft-core polymer model yields an effective colloidal pair potential
that decreases in strength and range with increasing $\alpha$
and~$\varepsilon$. In addition, this potential exhibits a repulsive
barrier for a separation $D \gtrsim R_{\rm G}$, owing to many-body
effects that arise from the mutual repulsion between polymers.  By
contrast, there is good agreement between the additive hard-sphere
mixture and the Gaussian polymer model, in particular for $\varepsilon
\approx 2 k_{\rm{B}}T$ and $\alpha \lesssim 1.5$.  Interestingly, these
are precisely the values that were found to reasonably represent dilute
and semidilute self-avoiding random walk
polymers~\cite{louis00a,louis00b,likos01}.

\section{Summary and conclusions}

We have presented a detailed description of the generalized geometric
cluster algorithm for continuum fluids introduced in Ref.~\cite{geomc},
which is a generalization of the work by Dress and
Krauth~\cite{dress95}.  In order to emphasize the connection with the
Swendsen--Wang algorithm for lattice models we have derived a
multiple-cluster variant of the GCA, in which a particle configuration
is decomposed into clusters that can be independently point-reflected
with respect to a given pivot point, without affecting the remainder of
the system. This algorithm establishes that it is generally possible to
identify such clusters and to devise a rejection-free Monte Carlo
scheme, independent of the detailed nature of the pair potentials
between particles. No restrictions are imposed upon the number of
species involved, but ergodicity can only be maintained if not every
cluster contains all particles of a given species. For monodisperse
hard-sphere fluids this implies a threshold packing fraction
around~$0.23$, which can be increased to approximately~$0.34$ if the
pivot is placed in a biased manner. For interactions with a longer
range, the clusters will typically be larger and hence the maximum
accessible volume fraction decreases. The nonlocal character of the
particle translations permits the efficient decorrelation of fluid
mixtures that involve a strong size asymmetry between the components.
Unavoidably, the resulting dynamics have no physical interpretation, but
all thermodynamic equilibrium properties are identical to those sampled
by conventional algorithms. While cluster-size distributions
for the Lennard-Jones fluid indicate that the percolation
threshold and the critical point do not coincide, the algorithm
significantly accelerates canonical simulations at the critical point.

Two specific example applications have been discussed, both involving
depletion interactions in size-asymmetric binary mixtures. In the first
illustration, we assess the effect of attractive interactions between
unlike species on the pair correlation function of the larger
species. This system has been investigated earlier by means of the
hard-sphere GCA supplemented by an acceptance criterion, and is included
here to illustrate that it is possible to perform the calculation in a
rejection-free manner. In the second illustration, we calculate the
depletion interaction induced by a depletant that acts as a hard sphere
for the larger species but has a Gaussian interaction potential with
other depletant particles. It is demonstrated how such calculations can
be performed efficiently for relatively large size ratios and lead to
depletion potentials that interpolate between the well-known AO
potential and the depletion potential for binary hard-sphere mixtures. 

A variety of extensions to the GCA can be devised. In particular, the
application to nonspherical particles is straightforward, but may
require an additional (rotational) Monte Carlo move that permits an
efficient relaxation of the rotational degrees of freedom. Periodic
boundary conditions are an essential ingredient for the point-reflection
moves employed, but application to layered systems (i.e., with only a
periodicity in the $x$ and~$y$ directions) can be realized if the point
reflection is performed in a horizontal plane and the relaxation along
the $z$-coordinate is performed via local Monte Carlo moves.
For strongly aspherical particles, the overlap threshold---and hence the
range of accessible volume fractions---decreases significantly.

In summary, the generalized GCA offers a wide range of opportunities to
efficiently simulate fluid systems that were hitherto inaccessible to
computer simulations. In addition, it may well be possible to generalize
this algorithm to other situations in which cluster algorithms are
employed, such as quantum-mechanical systems.

\begin{acknowledgments}
  This material is based upon work supported by the National Science
  Foundation under CAREER Award No.\ DMR-0346914 and Grant No.\
  CTS-0120978 and by the U.S.  Department of Energy, Division of
  Materials Sciences under Award No.\ DEFG02-91ER45439, through the
  Frederick Seitz Materials Research Laboratory at the University of
  Illinois at Urbana-Champaign.
\end{acknowledgments}

%\bibliographystyle{apsrev} 
%\bibliography{journals,simu,colloids,electrolyte,crossover,misc,gcalong-notes}

\begin{thebibliography}{49}
\bibitem[{not({\natexlab{b}})}]{note-parameters}
\bibinfo{note}{There appears to be some confusion regarding the parameters
  $\lambda$ (denoted $\kappa$ in the present work) and $\varepsilon$ employed
  in Ref.~\protect\cite{malherbe99}. The value $\lambda\sigma_{12} = 4$ (i.e.,
  $\lambda\sigma_{11}=4/3$) as quoted in the caption of Fig.~4 in
  Ref.~\protect\cite{malherbe99} is not correct (cf.\ also the caption of
  Fig.~3 in Ref.~\protect\cite{louis01b}, where a value $\lambda\sigma_{11}=4$
  is used, as in the present work). In addition, Ref.~\cite{malherbe99} quotes
  a value~$\beta\varepsilon/\sigma_{12}=1.6$. This interaction strength, which
  is three times stronger than in our case, actually does not yield the
  depicted~$g(r)$. Note that Ref.~\cite{louis01b}, on the other hand, quotes
  $\beta\varepsilon = -0.32$ (i.e., $1.6/5$) which implies that a
  factor~$\sigma_{\rm 22}$ has been omitted in their definition of the Yukawa
  potential. Furthermore, this definition contains a spurious factor~$\beta$.}
\end{thebibliography}

\end{document}